\documentclass[a4paper,11pt]{article}
\usepackage{amsfonts}
\usepackage{latexsym}
\usepackage{amsmath,amssymb}
\usepackage{mathrsfs}
\usepackage{cite}

\def\bbc{{\Bbb C}}
\def\bbr{{\Bbb R}}
\def\bbz{{\Bbb Z}}

\newtheorem{rem}{Remark}

\newtheorem{theo}{Theorem}

\newtheorem{defi}{Definition}
\newtheorem{ex}{Example}

\def\tr{\mathrm{tr\,}}

\def\ad{\mathrm{ad\,}}

\def\diag{\mathrm{diag\,}}
\def\rmi{\mathrm{i}}

\def\rmd{\mathrm{d\,}}

\begin{document}
\title{A Generic Nonlinear Evolution Equation\\ of Magnetic Type I. Reductions}
\author{Tihomir Valchev \\
\small Institute of Mathematics and Informatics,\\
\small Bulgarian Academy of Sciences,\\
\small Acad. Georgi Bonchev Str., Block 8, 1113 Sofia, Bulgaria,\\
\small tiv@math.bas.bg}
\date{}
\maketitle

\begin{abstract}
The present preprint is dedicated to a nonlinear evolution equation that generalizes the classical
Heisenberg ferromagnet equation in certain way. That generalization is completely integrable and has
a linear bundle Lax pair in pole gauge related to the special linear algebra. A few reductions of the
generic matrix equation are considered.
\\[0.2cm]
\end{abstract}

\section{Introduction}\label{intro}

In the present manuscript we shall consider a $1+1$-dimensional multicomponent nonlinear partial differential equation
that generalizes the Heisenberg ferromagnet equation 
\begin{equation}
\rmi S_t = \frac{1}{2} [S,S_{xx}],\qquad S(x,t)\in\mathfrak{sl}(2,\bbc),
\qquad S^2 = I.
\label{hf_0}
\end{equation}
Above "$\rmi$" denotes the imaginary unit, $I$ stands for the identity matrix of appropriate dimension, $[.,.]$ is the 
usual matrix commutator and the subscripts mean partial derivatives with respect to the independent variables. Ever
since the complete integrability of (\ref{hf_0}) was proved by Zakharov and Takhtadzhyan \cite{zakhtakh}, plenty of
integrable analogues or generalizations have been introduced. Basically, those nonlinear evolution equations (NEEs)
can be grouped into two different categories. The first one includes multicomponent analogues, i.e. the number of
dynamical fields (dependent variables) is increased. Two representative examples are the Heisenberg ferromagnet
equation related to an arbitrary simple Lie algebra \cite{forkul} and multicomponent Landau-Lifshitz equation
\cite{golsok}.

The second category consists in multidimensional NEEs, e.g. the $2+1$-dimensional Ishimori equation \cite{ishim}
\[\begin{split}
& \mathbf{S}_t = \mathbf{S}\times (\mathbf{S}_{xx} + \mathbf{S}_{yy}) + u_x\mathbf{S}_y
+ u_y \mathbf{S}_x,\\
& u_{xx} - u_{yy} =  - 2\mathbf{S} . (\mathbf{S}_x \times \mathbf{S}_y),
\end{split}\]
where $\mathbf{S}$ is a $3$-vector of unit length and $u$ is some real-valued function. Another similar example
is given by the $2+1$-dimensional Myrzakulov I equation \cite{mmnl}. 

The main object of study in this preprint is the following $1+1$-dimensional matrix NEE
\begin{equation}
\rmi S_t = a [S,S_{xx}] + \frac{3a}{2} \left(S^2S_{x}S\right)_x - b\left[S^2, S_{x}\right]_x ,\qquad
S(x,t)\in\mathfrak{sl}(n,\bbc),\quad a,b\in\bbc
\label{ghf_0}\end{equation}
that represents a natural generalization of the equation (\ref{hf_0}) and seems to be a novel one. For that reason
(\ref{ghf_0}) will further be referred to as generalized Heisenberg ferromagnet equation. The manuscript is
structured  as follows. In section \ref{genericghf} we shall prove that the equation (\ref{ghf_0}) has a Lax representation
provided $S(x,t)$ obeys some additional algebraic constraint. Section \ref{gauge trans} is dedicated to finding a
canonical representation of the Lax pair of (\ref{ghf_0}). In section \ref{reduction} we shall consider some local
and nonlocal reductions of the above matrix equation. Last section contains our concluding remarks.

\section{Generic Generalized Heisenberg Ferromagnet Equation}\label{genericghf}

In this section we shall propose a generalization of the Heisenberg ferromagnet equation that is
integrable in the sense of inverse scattering transform. This generalized Heisenberg ferromagnet
equation (GHF) has a zero curvature representation with a Lax pair generalizing in a natural way
the corresponding pair for the equation (\ref{hf_0}).

Let us consider the $1+1$-dimensional matrix NEE
\begin{equation}
\rmi S_t = a [S,S_{xx}] + \frac{3a}{2} \left(S^2S_{x}S\right)_x - b\left[S^2, S_{x}\right]_x ,
\qquad a,b\in\bbc
\label{ghf}\end{equation}
for the function $S:\bbr^2\to\mathfrak{sl}(n,\bbc)$ that is smooth enough almost everywhere. We shall require 
that the traceless complex $n\times n$-matrix $S(x,t)$ satisfies the following polynomial equation
\begin{equation}
S^3 = S .
\label{s_constr}
\end{equation}

\begin{rem}
The constraint (\ref{s_constr}) ensures that $S(x,t)$ has a constant spectrum. Indeed, it is easily seen from
(\ref{s_constr}) that $S(x,t)$ is a diagonalizable matrix with eigenvalues $0$ and $\pm 1$. Since $S(x,t)$ is 
a traceless matrix, the eigenvalues $1$ and $-1$ have the same multiplicities denoted by $r$. Moreover, $S^2(x,t)$
is an idempotent matrix of rank $2r$. In the special case when $S^2(x,t)$ is a nonsingular matrix (\ref{s_constr})
we immediately deduce that $S^2(x,t) = I$ (we remind that $I$ is the identity matrix).
\label{s_rem}\end{rem} 

\begin{rem}
From the previous remark it is straightforward that the adjoint operator $\ad_{S}(.):= [S, .]$ is
diagonalizable too and its spectrum consists of: $\pm 2$, $\pm 1$ and $0$ (with certain multiplicities).
Therefore $\ad_{S}$ obeys the relation
\begin{equation}
\ad_{S}^5 - 5\ad_{S}^3 + 4\ad_{S} = 0. 
\label{ads_chareq}\end{equation}	
\end{rem}

The equation (\ref{ghf}) contains the Heisenberg ferromagnet equation
\begin{equation}
\rmi S_t = \frac{1}{2}[S,S_{xx}], \qquad S(x,t)\in\mathfrak{sl}(n,\bbc),
\qquad S^2 = I
\label{hf}\end{equation}
as a particular case. In order to see this, let us require that $S(x,t)$ is an nonsingular matrix, i.e.
we have that $S^2(x,t) = I$. Taking into account that
\[S\,S_x = - S_x\, S\]
and setting $a=2$ in (\ref{ghf}), one immediately derives the equation (\ref{hf}).

Another special case of GHF that can be found in literature \cite{gmv,varna2020} is 
\begin{equation}
\rmi S_t + \left[S^2, S_{x}\right]_x = 0,
\label{ghf_gmv}\end{equation}
that is derived from (\ref{ghf}) when setting $a=0$. Such "reduction" is tightly related to a Hermitian
symmetric space of the type $\mathrm{SU}(n)\slash\mathrm{S}(\mathrm{U}(m)\times\mathrm{U}(n-m))$ as discussed
in \cite{gmv,varna2020}.
		
It is known that both (\ref{hf}) and (\ref{ghf_gmv}) admit zero curvature representation and one can apply
inverse scattering transform to solve them \cite{forkul,gmv,book,varna2020}. Similarly, for GHF the following
result holds true.

\begin{theo}
The generalized Heisenberg ferromagnet equation (\ref{ghf}) possesses a zero curvature representation.
\end{theo} 

\noindent {\bf Proof:} In order to prove this, let us consider the Lax pair:	
\begin{eqnarray}
L(\lambda) & = & \rmi\partial_x - \lambda S(x,t),\label{lax1}\\
A(\lambda) & = & \rmi\partial_t + \lambda V_1(x,t) + \lambda^2 V_2(x,t),\label{lax2}
\end{eqnarray}
where $\lambda\in\bbc$ is spectral parameter and $S(x,t)\in\mathfrak{sl}(n,\bbc)$ obeys the constraint
(\ref{s_constr}). The compatibility condition
\[ [L(\lambda),A(\lambda)] = 0 \]
of (\ref{lax1}) and (\ref{lax2}) must hold identically in the spectral parameter. Therefore it gives rise
to the following recurrence relations: 
\begin{eqnarray}
&&[S,V_{2}] = 0, \label{l3}\\
&&\rmi V_{2,x}  - [S, V_{1}] = 0, \label{l2}\\
&& V_{1,x} + S_t = 0 \label{l1}
\end{eqnarray}
that allow one to express $V_1$ and $V_2$ through $S$ (the potential function) and its $x$-derivative.
In view of (\ref{l3}) and the constraint (\ref{s_constr}) we can pick up $V_2$ to be
\begin{equation}
V_2 = a S + b\left(\frac{2r I}{n} - S^2\right), 
\label{a2}\end{equation}
where $a$ and $b$ are some complex valued functions, and $2r = \tr S^2$ (see Remark \ref{s_rem}).
The functions $a$ and $b$ are to be determined from the second recurrence relation. After substituting
(\ref{a2}) into (\ref{l2}), we obtain
\begin{equation}
a_x S + aS_x + b_x\left(\frac{2r I}{n} - S^2\right) - b\left(S^2\right)_x 
+ \rmi [S, V_{1}] = 0. \label{l2a}
\end{equation}
At this point we shall make use of a splitting of the coefficient $V_1$ that respects the structure of (\ref{l2a}).
More specifically, let us introduce a splitting of the form:
\[V_1 = V_1^{\mathrm{c}} + V_1^{\mathrm{r}},\qquad\left[S, V_1^{\mathrm{c}}\right] = 0.\]
For such splitting to be uniquely determined, we shall also require that
\[P_{S}V_1^{\mathrm{c}} = 0, \qquad P_{S}V_1^{\mathrm{r}} = V_1^{\mathrm{r}},\]
where the projector $P_S$ is defined through the relations
\[P_{S}:=\ad^{-1}_S\ad_{S}, \qquad \ad_{S}^{-1}:= \frac{1}{4}\left(5\ad_{S} - \ad_{S}^3\right).\]
The choice of $\ad_{S}^{-1}$ made above follows directly from the minimal characteristic equation for the adjoint
operator, see (\ref{ads_chareq}). 

Let us now apply the projector $P_S$ to the equation (\ref{l2a}). Taking into account the relations
\begin{eqnarray*}
P_{S}[S,X] &=& [S,X],\qquad \forall\, X\in\mathfrak{sl}(n,\bbc),\\
P_{S}S_x &=& S_x,\qquad P_{S}\left(S^{2}\right)_x = \left(S^{2}\right)_x,
\end{eqnarray*}
we can convince ourselves that (\ref{l2a}) splits into:
\begin{eqnarray}
a_x S + b_x\left(\frac{2r I}{n} - S^2\right) = 0,\label{l2b}\\
aS_x - b\left(S^2\right)_x + \rmi [S, V_{1}] = 0.\label{l2c}
\end{eqnarray}
After multiplying both hand sides of the former equation by $S$ and evaluating the trace,
we get
\[a_x = 0\qquad\Rightarrow\qquad a = a(t).\] 
Similarly, after multiplying both hand sides of (\ref{l2b}) by $S^2$ and evaluating the trace,
we have
\[b_x\left(\frac{2r}{n} - 1\right) = 0\qquad\Rightarrow\qquad b_x = 0\quad\cup\quad 2r = n.\]
The case when $2r = n$ means that $S^2 = I$ so this is effectively equivalent to
\[b=0\qquad\Rightarrow\qquad b_x = 0.\]
This is why both cases lead to the conclusion that $b$ is a scalar function of time only.

Equation (\ref{l2c}) allows us to find $V^{\mathrm{r}}_1$ in terms of $S$ and $S_x$ by inverting the commutator
in (\ref{l2c}). The result reads: 
\begin{equation}
V_1^{\mathrm{r}} = \rmi a\left([S,S_x] + \frac{3}{2}S^2S_xS\right) + \rmi b\left[S_x,S^2\right].
\label{a1r}\end{equation}
Now we substitute (\ref{a1r}) into (\ref{l1}) to obtain
\begin{equation}
\rmi V_{1,x}^{\mathrm{c}} - a \left([S,S_{x}] + \frac{3}{2}S^2S_xS\right)_x + b\left[S^2, S_x\right]_x + \rmi S_{t} = 0.
\label{l1det}\end{equation}
In view of the equalities
\begin{eqnarray*}
P_{S} \left[S, S_x\right]_x  &=& \left[S, S_x\right]_x,\qquad P_{S} \left[S^2, S_x\right]_x = \left[S^2, S_x\right]_x P_{S},\\ \left(S^2S_xS\right)_x  &=& \left(S^2S_xS\right)_x,\qquad P_{S} S_{t} = S_{t}
\end{eqnarray*}
the equation (\ref{l1det}) splits into
\[(\mathrm{id} - P_{S})V_{1,x}^{\mathrm{c}} = 0,\qquad
\rmi S_t + \rmi P_{S}(V_{1,x}^{\mathrm{c}}) = a \left([S,S_{x}] + \frac{3}{2}S^2S_xS\right)_x
- b \left[S^2, S_x\right]_x, \]
where "$\mathrm{id}$" denotes the identity operator. Setting $V_1^{\mathrm{c}} = 0$ or any constant $S$-commuting
matrix, we immediately derive the following NEE
\[\rmi S_t = a [S,S_{xx}] + \frac{3a}{2} \left(S^2S_{x}S\right)_x - b\left[S^2, S_{x}\right]_x .\]
This way we have proved that (\ref{ghf}) represents the zero curvature condition of the Lax pair
\begin{eqnarray}
L(\lambda) & = & \rmi\partial_x - \lambda S(x,t),\qquad S^3 = S,\label{lax1a}\\
A(\lambda) & = & \rmi\partial_t + \lambda V_1(x,t) + \lambda^2 V_2(x,t),\qquad 
V_2 = aS + b\left(\frac{2r}{n}I - S^2\right),\label{lax2a}\\
V_1 & = & \rmi a\left([S,S_x] + \frac{3}{2}S^2S_xS\right) + \rmi b\left[S_x,S^2\right]
\label{lax3a}
\end{eqnarray}
even when the coefficients $a$ and $b$ are not restricted to be complex numbers but could be
any complex-valued smooth functions of time. \qquad $\blacksquare$

In the following section we shall focus on the Lax pair (\ref{lax1a})--(\ref{lax3a}) and discuss one of its
essential properties.

\section{Gauge Transformations and Gauge Equivalent Lax Pairs}\label{gauge trans}

Here we shall get acquainted with the canonical representation of the Lax pair (\ref{lax1a})--(\ref{lax3a}).
This representation underlies the formalism of direct scattering problem used to study and solve a NEE
integrable through inverse scattering transform.

We shall start by briefly reminding the notion of gauge transformations and gauge equivalent Lax pairs. We
refer the reader to \cite{book} where they can find more details. Let us introduce the Lax operators in general
form:
\begin{equation}
L(\lambda) = \rmi\partial_x + U(x,t,\lambda),\qquad A(\lambda) = \rmi\partial_t + V(x,t,\lambda),
\label{la_gen}
\end{equation}
where $U(x,t,\lambda), V(x,t,\lambda)\in\mathfrak{sl}(n,\bbc)$ depend on the spectral parameter $\lambda$ in
some unspecified way. Let $\Psi$ be an arbitrary fundamental set of solutions of the overdetermined system
\begin{eqnarray}
&&L(\lambda)\Psi(x,t,\lambda) = 0, \label{sp_problem}\\
&&A(\lambda)\Psi(x,t,\lambda) = 0.
\label{seclinpr}
\end{eqnarray}
We shall denote by $\mathbb{S}$ the set of all fundamental sets of solutions (fundamental solutions for short)
to the linear problems (\ref{sp_problem}) and (\ref{seclinpr}). The formal integrability condition of (\ref{sp_problem})
and (\ref{seclinpr}) gives rise to the following equation for $U$ and $V$:
\begin{equation}
\rmi U_t - \rmi V_x - [U,V] = 0.
\label{zcc_gen}\end{equation}
That equation has a large number of symmetries. Indeed, let us introduce the transformation
\begin{equation}
\mathbb{S}\to \tilde{\mathbb{S}} := \mathcal{G}\,\mathbb{S} = \{\mathcal{G}\,\Psi,\quad \Psi\in\mathbb{S}\},
\label{gauge_tr}\end{equation}
where $\mathcal{G}(x,t)$ is some $n\times n$-matrix with unit determinant. Then the Lax pair (\ref{la_gen}) is transformed
under (\ref{gauge_tr}) according to the rule:
\begin{eqnarray*}
L(\lambda)\to \tilde{L}(\lambda) &=& \mathcal{G}(x,t)L(\lambda)\left(\mathcal{G}(x,t)\right)^{-1},\\
A(\lambda)\to \tilde{A}(\lambda) &=& \mathcal{G}(x,t)A(\lambda)\left(\mathcal{G}(x,t)\right)^{-1}.
\end{eqnarray*}
The transformed differential operators can be written down as
\[\tilde{L}(\lambda) = \rmi\partial_x + \tilde{U}(x,t,\lambda),\qquad
\tilde{A}(\lambda) = \rmi\partial_t + \tilde{V}(x,t,\lambda),\]
where 
\begin{eqnarray}
\tilde{U}(x,t,\lambda) &=& \rmi\mathcal{G}(x,t)\partial_x\left(\mathcal{G}(x,t)\right)^{-1}
+ \mathcal{G}(x,t)U(x,t,\lambda)\left(\mathcal{G}(x,t)\right)^{-1},\label{U_gauge_tr}\\
\tilde{V}(x,t,\lambda) &=& \rmi\mathcal{G}(x,t)\partial_t\left(\mathcal{G}(x,t)\right)^{-1}
+ \mathcal{G}(x,t)V(x,t,\lambda)\left(\mathcal{G}(x,t)\right)^{-1}.\label{V_gauge_tr}
\end{eqnarray}
In the rather important polynomial bundle case, i.e. when $U(x,t,\lambda) = \sum^{M}_{k=0}U_k(x,t)\lambda^k$ and
$V(x,t,\lambda) = \sum^{N}_{l=0}V_l(x,t)\lambda^{l}$, the transformed matrices $\tilde{U}(x,t,\lambda)$ and
$\tilde{V}(x,t,\lambda)$ are polynomials whose coefficients are given by the relations:
\begin{eqnarray}
\tilde{U}_0 &=& \rmi\mathcal{G}\partial_x\mathcal{G}^{-1} + \mathcal{G}U_0\mathcal{G}^{-1},\qquad
\tilde{U}_p = \mathcal{G}U_p\mathcal{G}^{-1},\quad p=1,\ldots,M, \label{U_gauge_tr_a}\\
\tilde{V}_0 &=& \rmi\mathcal{G}\partial_t\mathcal{G}^{-1} + \mathcal{G}V_0\mathcal{G}^{-1},\qquad
\tilde{V}_q = \mathcal{G}V_q\mathcal{G}^{-1},\quad q = 1,\ldots,N.\label{V_gauge_tr_a}
\end{eqnarray}
It is straightforward to see that (\ref{zcc_gen}) is invariant under (\ref{U_gauge_tr}) and (\ref{V_gauge_tr}),
i.e. we have 
\[\rmi \tilde{U}_t - \rmi \tilde{V}_x - [\tilde{U},\tilde{V}] = 0.\]
		
\begin{defi}
The transformations just introduced are called gauge transformations and they constitute an infinite dimensional
Lie group. The form of the pair $(U,V)$ is called gauge. Two Lax pairs are called gauge equivalent if those are
connected through some gauge transformation. 
\end{defi}
		
Now let's turn back to the linear bundles under consideration, i.e. the Lax pair is in the form (\ref{lax1}) and
(\ref{lax2}). The following theorem holds true.
\begin{theo}
The scattering operator (\ref{lax1a}) is gauge equivalent to
\begin{equation}
\tilde{L}(\lambda) = \rmi\partial_x + Q(x,t) - \lambda \Sigma,\qquad
\Sigma = \diag (I_r, 0, -I_r), 
\label{L_canon}\end{equation}
where $Q(x,t)$ is some off-diagonal traceless matrix with complex entries, $I_r$ stands for the identity matrix of
dimension $r$ ($r = \tr S^2/2$). 
\end{theo}

\noindent{\bf Proof:}	
As we discussed in the previous section, see Remark \ref{s_rem}, there exists a gauge transformation
$\mathcal{G}_0(x,t)$ that diagonalizes the matrix $S(x,t)$. More specifically, we have that
\[\mathcal{G}_0(x,t)S(x,t)\left(\mathcal{G}_0(x,t)\right)^{-1} = \Sigma, \qquad
\Sigma = \diag (I_r, 0, -I_r)\]
and (\ref{lax1a})--(\ref{lax3a}) is transformed into the Lax pair:
\begin{eqnarray}
\overline{L}(\lambda) &=& \rmi\partial_x + \overline{U}_0(x,t) - \lambda\Sigma ,\label{lax1b}\\
\overline{A}(\lambda) &=& \rmi\partial_t + \overline{V}_0(x,t) + \lambda\overline{V}_1(x,t)
+ \lambda^2 \varXi , \label{lax2b}
\end{eqnarray}
where 
\begin{eqnarray*}
\overline{U}_0(x,t) &=& \rmi\mathcal{G}_0(x,t)\partial_x\left(\mathcal{G}_0(x,t)\right)^{-1},\qquad
\overline{V}_0(x,t) = \rmi\mathcal{G}_0(x,t)\partial_t\left(\mathcal{G}_0(x,t)\right)^{-1},\\
\overline{V}_1(x,t) &=& \mathcal{G}_0(x,t)V_1(x,t)\left(\mathcal{G}_0(x,t)\right)^{-1},\qquad
\varXi = a\Sigma + b\left(\frac{2r}{n}I_n - \Sigma^2\right). 
\end{eqnarray*}
Let us denote by $\overline{u}_0(x,t)$ the diagonal part of the traceless matrix $\overline{U}_0(x,t)$. If
$\overline{u}_0(x,t)$ is zero then the theorem is already proved. If $\overline{u}_0(x,t)\neq 0$, we apply
another gauge transformation $\mathcal{G}_1(x,t)$ sought in the form:
\begin{equation}
\mathcal{G}_1(x,t) = \exp\left(\rmi F(x,t)\right).
\label{g1}\end{equation}
Above $F(x,t)$ is some diagonal traceless $n\times n$-matrix with complex entries to be determined from the diagonal part
of $\overline{U}_0(x,t)$. The ansatz (\ref{g1}) ensures that the $\lambda$-leading terms of the Lax pair (\ref{lax1b}) and
(\ref{lax2b}) remain intact under the gauge transformation $\mathcal{G}_1(x,t)$.

After applying (\ref{g1}) onto (\ref{lax1b}) and (\ref{lax2b}), we obtain the Lax pair
\begin{eqnarray}
\tilde{L}(\lambda) &=& \rmi\partial_x + \tilde{U}_0(x,t) - \lambda\Sigma ,\label{lax1c}\\
\tilde{A}(\lambda) &=& \rmi\partial_t + \tilde{V}_0(x,t) + \lambda\tilde{V}_1(x,t)
+ \lambda^2 \varXi , \label{lax2c}
\end{eqnarray}
where
\begin{eqnarray*}
\tilde{U}_0(x,t) &=& F_x(x,t) + \exp\left(\rmi F(x,t)\right)\overline{U}_0(x,t)\exp\left(-\rmi F(x,t)\right),\\
\tilde{V}_0(x,t) &=& F_t(x,t) + \exp\left(\rmi F(x,t)\right)\overline{V}_0(x,t)\exp\left(-\rmi F(x,t)\right),\\
\tilde{V}_1(x,t) &=& \exp\left(\rmi F(x,t)\right)\overline{V}_1(x,t)\exp\left(-\rmi F(x,t)\right).
\end{eqnarray*}
Now we require that the diagonal part of $\tilde{U}_0(x,t)$ vanishes. This leads to the following simple differential
equation for $F$
\[F_x + \overline{u}_0(x,t) = 0,\]
that is immediately solved to give
\begin{equation}
F(x,t) = - \int\overline{u}_0(x,t)\rmd x.
\label{gauge2_dex}
\end{equation}
After applying the gauge transform $\mathcal{G}_1\mathcal{G}_0$ to (\ref{lax1a}) and replacing
$\tilde{U}_0(x,t)$ with $Q(x,t)$ in (\ref{lax1c}), we derive a Lax operator of the canonical form
(\ref{L_canon}). This way we have proved that the Lax operators (\ref{lax1a}) and (\ref{L_canon}) are gauge equivalent.
\qquad $\blacksquare$

\begin{rem}
The above theorem just states the existence of a gauge transformation with prescribed properties. On the other hand,
it is evident that (\ref{gauge2_dex}) does not determine $F(x,t)$ uniquely --- it is up to some function of $t$.
In order to reduce this ambiguity, one could impose a stronger condition --- that both diagonal parts of
$\tilde{U}_0(x,t)$ and $\tilde{V}_0(x,t)$ vanish simultaneously. As a result, for $F(x,t)$ one obtains that 
\[F(x,t) = - \int\overline{u}_0(x,t)\rmd x + \mathrm{const} = - \int\overline{v}_0(x,t)\rmd t + \mathrm{const}' ,\]
where $\overline{v}_0$ stands for the diagonal part of $\overline{V}_0$.
\end{rem}

\section{Reductions of GHF}\label{reduction}

In this section we shall consider certain local and nonlocal reductions of the generic matrix NEE
(\ref{ghf}). This is why we shall briefly remind the notion of reductions and reduction group, referring
the reader to \cite{mikh1,mikh2,pla} where they can find more detailed explanations. 

Assume the linear problems (\ref{sp_problem}) and (\ref{seclinpr}) admit a finite symmetry group
$\mathrm{G}_{\rm R}$, i.e. if $\Psi$ is a fundamental solution to (\ref{sp_problem}) and (\ref{seclinpr}) then
\[\tilde{\Psi}(x,t,\lambda):= \mathrm{K}_g\left(\Psi(\kappa^{-1}_{g}(x,t),k^{-1}_g(\lambda))\right),
\qquad g\in\mathrm{G}_{\mathrm{R}}\]
gives another fundamental solution. Above $\kappa_g:\bbr^2\to\bbr^2$ is a smooth mapping, $k_g$
is a conformal mapping, and $\mathrm{K}_g$ is a group automorphism of $\mathrm{SL}(n,\bbc)$.
The invariance of (\ref{sp_problem}) and (\ref{seclinpr}) under the action of $\mathrm{G}_{\rm R}$ effectively
decreases the number of independent entries in $U(x,t,\lambda)$ and $V(x,t,\lambda)$. As a result, the NEEs
representing the zero curvature condition (\ref{zcc_gen}) of the Lax pair (\ref{la_gen}) is "reduced".
This motivates the following definition.
	
\begin{defi}
The group $\mathrm{G}_{\mathrm{R}}$ is called reduction group while the $\mathrm{G}_{\mathrm{R}}$-action on
the set of fundamental solutions $\mathbb{S}$ and on the Lax operators bears the name reduction. In the case
when $\mathrm{G}_{\mathrm{R}}$ does not transform the variables $x$ and $t$, the reduction is termed local,
otherwise it is called nonlocal. 
\end{defi}

Generally speaking a discrete symmetry of the linear problem (\ref{sp_problem}) does not necessarily yield a
reduction of a NEE whose scattering operator is $L(\lambda)$. This is why the invariance of the second linear
problem is essential here. Let us apply those general concepts to the NEE under consideration. We shall start
with a simple example of local reduction induced by the group $\bbz_2$.

\begin{ex}{\bf (Cartan reduction)}

Assume that the linear problems (\ref{sp_problem}) and (\ref{seclinpr}) remain invariant under the $\bbz_2$-action  
\begin{equation}
\Psi(x,t,\lambda)\to \tilde{\Psi}(x,t,\lambda) = C\Psi(x,t,-\lambda)C^{-1},\qquad
C = \diag(-I_m,I_{n-m}).
\label{cartanred}\end{equation}
It is easily seen that the invariance of both linear problems under (\ref{cartanred}) gives rise to
the symmetry conditions
\begin{eqnarray}
CL(-\lambda)C^{-1} &=& L(\lambda) \quad\Leftrightarrow\quad CU(x,t,-\lambda)C^{-1} = U(x,t,\lambda), 
\label{cartanred1}\\
CA(-\lambda)C^{-1} &=& A(\lambda) \quad\Leftrightarrow\quad CV(x,t,-\lambda)C^{-1} = V(x,t,\lambda) .
\label{cartanred2}
\end{eqnarray}
For polynomial bundles (\ref{cartanred1}) and (\ref{cartanred2}) split into 
\begin{eqnarray*}
CU_k(x,t)C^{-1} &=& (-1)^{k}U_k(x,t),\qquad k = 0,\ldots, M,\\
CV_l(x,t)C^{-1} &=& (-1)^{l}V_l(x,t), \qquad l = 0,\ldots, N,
\end{eqnarray*}
where $U(x,t,\lambda) = \sum^{M}_{k=0}U_k(x,t)\lambda^k$ and $V(x,t,\lambda) = \sum^{N}_{l=0}V_l(x,t)\lambda^{l}$.

The matrix $C$ determines the adjoint action of the Cartan involution, defining the Hermitian symmetric space
$\mathrm{SU}(n)\slash \mathrm{S}(\mathrm{U}(m)\times\mathrm{U}(n-m))$, hence the name Cartan reduction for (\ref{cartanred}). 
The afore-mentioned adjoint action induces a $\bbz_2$-grading in the special linear algebra as follows:
\begin{eqnarray*}
\mathfrak{sl}(n,\bbc) &=& \mathfrak{sl}^0(n,\bbc) + \mathfrak{sl}^1(n,\bbc),\\
\mathfrak{sl}^{\varsigma}(n,\bbc) &=& \{X\in \mathfrak{sl}(n,\bbc);\ CXC^{-1} = (-1)^{\varsigma}X\},
\qquad\varsigma = 0,1.
\end{eqnarray*}
It is straightforward to see that for polynomial bundles we have
\begin{eqnarray*}
U_k(x,t)&\in&\mathfrak{sl}^{\varsigma}(n,\bbc),\qquad k\equiv \varsigma (\mathrm{mod}\, 2),\\
V_l(x,t)&\in&\mathfrak{sl}^{\varsigma}(n,\bbc),\qquad l\equiv \varsigma (\mathrm{mod}\, 2) .
\end{eqnarray*}
In view of the explicit form of $C$ the matrix coefficients of the Lax pair acquire block structure as given below:
\begin{eqnarray*}
U_k(x,t) = \left(\begin{array}{cc}
\star & 0 \\ 0 & \star
\end{array}\right),\quad k\equiv 0 (\mathrm{mod}\, 2), \qquad V_l(x,t) = \left(\begin{array}{cc}
\star & 0 \\ 0 & \star
\end{array}\right),\quad l\equiv 0 (\mathrm{mod}\, 2),\\
U_k(x,t) = \left(\begin{array}{cc}
0 &	\star \\ \star & 0
\end{array}\right),\quad k\equiv 1 (\mathrm{mod}\, 2), \qquad V_l(x,t) = \left(\begin{array}{cc}
0 &	\star \\ \star & 0
\end{array}\right),\quad l\equiv 1 (\mathrm{mod}\, 2).
\end{eqnarray*}
In this case it is said that the NEE derived from (\ref{zcc_gen}) is related to the symmetric
space $\mathrm{SU}(n)\slash \mathrm{S}(\mathrm{U}(m)\times\mathrm{U}(n-m))$. 

For the case of the linear bundle Lax pair (\ref{lax1a})--(\ref{lax3a}) the $\bbz_2$-reduction
(\ref{cartanred}) implies that $a\equiv 0$. Therefore GHF simplifies to the equation (\ref{ghf_gmv}).
It is natural to use the representation 
\begin{equation}
S(x,t) = \left(\begin{array}{cc}
0 & \mathbf{u}^T(x,t) \\ \mathbf{v}(x,t) & 0
\end{array}\right),
\label{s_blocks}\end{equation}
where the superscript $T$ means matrix transposition. Then (\ref{ghf_gmv}) can be written down in terms of
the complex $(n-m)\times m$-matrices $\mathbf{u}(x,t)$ and $\mathbf{v}(x,t)$ as follows:
\begin{eqnarray*}
\rmi\mathbf{u}_t + \left(\mathbf{u}_x\mathbf{v}^T\mathbf{u} - \mathbf{u}\mathbf{v}^T\mathbf{u}_x\right)_x &= 0,\\
\rmi\mathbf{v}_t - \left(\mathbf{v}_x\mathbf{u}^T\mathbf{v} - \mathbf{v}\mathbf{u}^T\mathbf{v}_x\right)_x &= 0 .
\end{eqnarray*}
Due to (\ref{s_constr}) the matrix solutions of the above system are interrelated through the cubic constraints
\[\mathbf{u}\mathbf{v}^T\mathbf{u} = \mathbf{u},\qquad \mathbf{v}\mathbf{u}^T\mathbf{v} = \mathbf{v}.\]

\end{ex}

While in the example just discussed both independent variables $x$ and $t$ remained unchanged, this is not true
in the next one.

\begin{ex}{\bf (Pseudo-Hermitian reduction)}

Let us consider the following action of the group $\bbz_2$
\begin{equation}
\Psi(x,t,\lambda) \to \tilde{\Psi}(x,t,\lambda) = \mathcal{E}\left[\Psi^{\dag}(\varepsilon x,t,\varepsilon\lambda^*)\right]^{-1}\mathcal{E}^{-1},\qquad \varepsilon^2 = 1,
\label{hermit_red}
\end{equation}
where $\mathcal{E}$ is a diagonal matrix with $\pm 1$ on its diagonal, $\dag$ stands for
Hermitian conjugation, and $*$ means complex conjugation. The invariance of the linear problems
(\ref{sp_problem}) and (\ref{seclinpr}) under (\ref{hermit_red}) now leads to the conditions:
\begin{eqnarray}
\mathcal{E}U^{\dag}(\varepsilon x,t,\varepsilon\lambda^*)\mathcal{E}^{-1} & = & \varepsilon U(x,t,\lambda), 
\label{hermitred1}\\
\mathcal{E}V^{\dag}(\varepsilon x,t,\varepsilon\lambda^*)\mathcal{E}^{-1} & = & V(x,t,\lambda).
\label{hermitred2}
\end{eqnarray}
In the special case when $U(x,t,\lambda) = \sum^{M}_{k=0}U_k(x,t)\lambda^k$ and $V(x,t,\lambda) =
\sum^{N}_{l=0}V_l(x,t)\lambda^{l}$ the symmetry conditions (\ref{hermitred1}) and (\ref{hermitred2})
give rise to:
\begin{eqnarray*}
\mathcal{E}U^{\dag}_k(\varepsilon x,t)\mathcal{E}^{-1} & = & \varepsilon^{k-1} U_k(x,t), \qquad k=0,\ldots, M,\\
\mathcal{E}V^{\dag}_l(\varepsilon x,t)\mathcal{E}^{-1} & = & \varepsilon^{l} V_l(x,t),\qquad l=0,\ldots, N.
\end{eqnarray*}

Evidently, (\ref{hermit_red}) represents a local $\bbz_2$ reduction if $\varepsilon = 1$ and a nonlocal one if
$\varepsilon = -1$. In the case of local reduction the matrix coefficients $U_k(x,t)$ and $V_l(x,t)$ can be interpreted
as generators of symmetry of a pseudo-Hermitian form given by the matrix $\mathcal{E}$. This is the reason why (\ref{hermit_red})
is termed pseudo-Hermitian reduction.

In the case of the Lax pair (\ref{lax1a})--(\ref{lax3a}) with the reduction (\ref{hermit_red}) imposed on it, GHF preserves
its form (\ref{ghf}) but now both constants $a$ and $b$ are real. If we combine (\ref{hermit_red}) with the Cartan reduction from the
previous example, we may easily deduce that the following interrelation for the potential $S(x,t)$ presented as
in (\ref{s_blocks}) holds true:
\[\mathbf{v}(x,t) = \mathcal{E}_n\mathbf{u}^*(\varepsilon x,t)\mathcal{E}_m.\]
Above $\mathcal{E}_{m}$ and $\mathcal{E}_{n}$ are diagonal matrices of dimension $m$ and $n$ respectively with
$\pm 1$ on their diagonals so that we have $\mathcal{E} = \diag(\mathcal{E}_{m}, \mathcal{E}_{n})$.
Then the matrix-valued function $\mathbf{u}$ satisfies the equation
\[\rmi\mathbf{u}_t + \left[\mathbf{u}_x(x,t)\mathcal{E}_m\mathbf{u}^{\dag}(\varepsilon x,t)\mathcal{E}_n\mathbf{u}(x,t)
- \mathbf{u}(x,t)\mathcal{E}_m\mathbf{u}^{\dag}(\varepsilon x,t)\mathcal{E}_n\mathbf{u}_x(x,t)\right]_x = 0\]
along with the constraint
\[\mathbf{u}(x,t)\mathcal{E}_{m}\mathbf{u}^{\dag}(\varepsilon x,t)\mathcal{E}_n\mathbf{u}(x,t)
= \mathbf{u}(x,t).\]

\end{ex}

We have considered so far Lax pairs with coefficients in the Lie algebra $\mathfrak{sl}(n,\bbc)$. It is a well-known
fact that any of the classical series of simple Lie algebras can be extracted from the special linear algebra by imposing
an appropriate $\bbz_2$ reduction as it will be demonstrated below.

\begin{ex}{\bf (Orthogonal/symplectic reduction)}
	
Let us introduce the $\bbz_2$-reduction:
\begin{equation}
\Psi(x,t,\lambda)\to \tilde{\Psi}(x,t,\lambda) = G^{-1}\left[\Psi^{T}(x,t,\lambda)\right]^{-1}G.
\label{orthsympl}\end{equation}
Although $G$ could be any constant $n\times n$-matrix to obey $G^2 = \pm I_n$, we shall be primarily
interested in the following two cases: symmetric case $\left(G^T = G\right)$ and skew-symmetric case
$\left(G^T = - G\right)$. Without loss of generality we may set
\[G_{ij} = \left\{\begin{array}{ll}
\delta_{i\, n+1-j},  & G^T = G, \\
(-1)^{i-1}\delta_{i\, n+1-j},& G^T = - G, \quad n\equiv 0 (\mathrm{mod}\, 2).
\end{array}
\right.\]

The  invariance of (\ref{sp_problem}) and (\ref{seclinpr}) under (\ref{orthsympl}) is equivalent to
the conditions
\begin{eqnarray}
U^{T}(x,t,\lambda)G + GU(x,t,\lambda) = 0, \label{orthsumpl1}\\
V^{T}(x,t,\lambda)G + GV(x,t,\lambda) = 0.
\label{orthsumpl2}
\end{eqnarray}
For bundles of the form
\[U(x,t,\lambda) = \sum^{M}_{k=0}U_k(x,t)\lambda^k,\qquad V(x,t,\lambda) =
\sum^{N}_{l=0}V_l(x,t)\lambda^{l}\]
(\ref{orthsumpl1}) and (\ref{orthsumpl2}) split into
\begin{eqnarray}
U^{T}_k(x,t)G + G U_k(x,t) &=& 0, \qquad k=0,\ldots, M,\label{orthsumpl1a}\\
V^{T}_l(x,t)G + GV_l(x,t) &=& 0,\qquad l=0,\ldots, N.\label{orthsumpl2a}
\end{eqnarray}  	
The equations (\ref{orthsumpl1a}) and (\ref{orthsumpl2a}) have the form of relations defining the  
generators of symmetry of a nondegenerate bilinear form $G$. In the symmetric case $G$ can naturally be
interpreted as a metric so $U_k(x,t)$ and $V_l(x,t)$ belong to the orthogonal algebra
$\mathfrak{so}(n,\bbc)\subset\mathfrak{sl}(n,\bbc)$. Similarly, in the skew-symmetric case $G$
plays the role of a symplectic form so $U_k(x,t)$ and $V_l(x,t)$ are elements of the symplectic algebra $\mathfrak{sp}(n,\bbc)\subset\mathfrak{sl}(n,\bbc)$.

For the linear bundle (\ref{lax1a})--(\ref{lax3a}) the potential $S(x,t)$ belongs either to
$\mathfrak{so}(n,\bbc)$ (symmetric case) or to $\mathfrak{sp}(n,\bbc)$ (skew-symmetric case). It is not
hard to see that $S^2$ (or any even power of $S$) {\bf does not} belong to $\mathfrak{so}(n,\bbc)$ or to
$\mathfrak{sp}(n,\bbc)$ respectively. Therefore we have that $V_2 \propto S$ ($b\equiv 0$) and the NEE
(\ref{ghf}) simplifies to 
\[\rmi S_t = a [S,S_{xx}] + \frac{3a}{2} \left(S^2S_{x}S\right)_x,
\qquad a\in\bbc .\]

\end{ex}

\section{Conclusion}

We have considered the generic matrix NEE (\ref{ghf}) generalizing the classical Heisenberg ferromagnet equation
(\ref{hf_0}). This generalized Heisenberg ferromagnet equation has a Lax representation which is gauge equivalent
to a linear bundle of the canonical form (\ref{lax1c}) and (\ref{lax2c}). We have demonstrated that (\ref{ghf})
contains as a $\bbz_2$-reduction another NEE of magnetic type that is related to a Hermitian symmetric space of
the type $\mathrm{SU}(n)\slash\mathrm{S}(\mathrm{U}(m)\times\mathrm{U}(n-m))$. We have discussed pseudo-Hermitian
local and nonlocal reductions of the generic equation (\ref{ghf}). Reductions of the generalized Heisenberg ferromagnet
equation leading to a NEE related to orthogonal and symplectic algebras have been considered as well. Of course,
all this is far from being a complete list of all admissible reductions.

The generalized Heisenberg ferromagnet equation is the simplest nontrivial member of an integrable hierarchy which
should be described in terms of generating operators. It is still an open question how to achieve such description
in a satisfactory way.

Finding particular solutions to (\ref{ghf}) is another essential issue. It is envisaged to be done elsewhere.

\section{Acknowledgements}
The work has been supported by Grant DN 02-5 of Bulgarian Fund "Scientific Research".


\begin{thebibliography}{99.}
\bibitem{forkul}
Fordy, A., Kulish, P., Nonlinear Schr\"{o}dinger Equations and Simple Lie
Algebras, {\it Commun. Math. Phys.} {\bf 89} (1983) 427--443.
\bibitem{book}
Gerdjikov, G., Vilasi, G., Yanovski, A., {\it Integrable Hamiltonian Hierarchies
--- Spectral and Geometric Methods}, Springer, Heidelberg, 2008.
\bibitem{gmv}
Gerdjikov, G., Mikhailov, A., Valchev, T., Reductions of Integrable Equations
on A.III-type Symmetric Spaces, {\it J. Phys. A: Math. Theor.} {\bf 43} (2010) 434015.
\bibitem{golsok}
Golubchik, I. Z., Sokolov, V. V., Multicomponent Generalization of the Hierarchy
of the Landau-Lifshitz Equation, {\it Theor. Math. Phys.} {\bf 124}, n. 1 (2000) 909--917.
\bibitem{ishim}
Ishimori, Y., Multi-vortex Solutions of a Two-dimensional Nonlinear Wave Equation,
{\it Prog. Theor. Phys.} {\bf 72} (1984) 33--37.
\bibitem{mikh1}
Mikhailov, A., Reduction in the Integrable Systems. Reduction Groups, {\it Lett. JETF} {\bf 32}
(1979) 187--192.
\bibitem{mikh2}
Mikhailov, A., The Reduction Problem and Inverse Scattering Method, {\it Physica D} {\bf 3} (1981) 73--117.
\bibitem{mmnl}
Myrzakulov, R., Mamyrbekova, G., Nugmanova, G., Lakshmanan, M., Integrable (2+1) Dimensional 
Spin Model: Geometric and Gauge Equivalent Counterparts, Solitons and Coherent Structures,
{\it Phys. Lett. A} {\bf 233} (1997) 391--396.
\bibitem{pla}
Valchev, T., On Mikhailov's Reduction Group, {\it Phys. Lett. A} {\bf 379} (2015) 1877–1880.  
\bibitem{varna2020}
Valchev, T., Multicomponent Nonlinear Evolution Equations of the Heisenberg Ferromagnet Type.
Local Versus Nonlocal Reductions, In: {\it Geometry, Integrability and Quantization}, {\bf 22},
Eds.: I. Mladenov, V. Pulov and A. Yoshioka, Avangard Prima, Sofia, 2021, 274--285 (Proc. XXII-nd
International Conference "Geometry, Integrability and Quantization", June 8--13, 2020, Varna, Bulgaria),
arXiv: 2011.13437v1[nlin.SI].
\bibitem{zakhtakh}
Zakharov, V., Takhtadzhyan, L., Equivalence of the Nonlinear Schr\"odinger Equation and the
Equation of a Heisenberg Ferromagnet, {\it Theor. Math. Phys.} {\bf 38} (1979) 17-–23.
\end{thebibliography}
\end{document}